# Evolving Gene Regulatory Networks with Mobile DNA Mechanisms

Larry Bull and Andrew Adamatzky

*Abstract*— This paper uses a recently presented abstract, tuneable Boolean regulatory network model extended to consider aspects of mobile DNA, such as transposons. The significant role of mobile DNA in the evolution of natural systems is becoming increasingly clear. This paper shows how dynamically controlling network node connectivity and function via transposon-inspired mechanisms can be selected for in computational intelligence tasks to give improved performance. The designs of dynamical networks intended for implementation within the slime mould *Physarum polycephalum* and for the distributed control of a smart surface are considered.

## I. INTRODUCTION

A number of mobile DNA mechanisms exist through which changes in genomic structure can occur in ways other than copy errors, particularly via transposable elements (e.g., see [Craig et al., 2002] for an overview). Mobile genetic elements such as transposons are DNA sequences that may be either copied or removed and then inserted at a new position in the genome [McClintock, 1987]: retrotransposons use an intermediary RNA copy of themselves for "copying and pasting", whereas DNA transposons rely upon specific proteins for their "cutting and pasting" into new sites, respectively. The targeting of a new position ranges from the very specific, typically by exploiting sequence recognition proteins, to more or less arbitrary movement. These processes, insertion in particular, are often reliant upon proteins produced elsewhere within the genome. Transposons are found widely in both prokaryotes and eukaryotes, and they have been associated with many significant evolutionary innovations (e.g., see [Kazazian, 2004] for an overview).

Transposable elements can therefore change the behaviour of a given cell: insertion into a gene will typically disrupt its coding sequence, i.e., it will be mutated, insertion next to a gene may affect its subsequent regulation, e.g., the mobile element's regulatory sequence may take control of the gene, the act of excision can leave behind DNA fragments which cause a change in the sequence at that location, coding segments between transposons can be moved with them, etc. The effects of such movement can be beneficial or detrimental to a cell. Perhaps the most significant aspect of transposons is that these effects occur *during* the organism/cell's lifetime. That is, *such structural changes are made to a genome based upon the actions of its own regulatory processes in response to its internal and external environment*. Moreover, such changes can be inherited. Thus, as has recently been highlighted [Shapiro, 2011], genomes should be viewed as read-write systems with embedded change/search heuristics.

This paper extends initial studies which began consideration of the dynamic role of mobile DNA within regulatory network representations [Bull, 2012a; 2013]. In particular, an aspect of transposable elements within a genetic regulatory network (GRN) was explored using an extension of a well-known, simple GRN formalism – random Boolean networks (RBN) [Kauffman, 1969]. The RBN model was extended to include a form of structural dynamism to capture some aspects of mobile DNA during the cell life-cycle: gene connectivity could be varied based upon the current network/environment state. It was shown that such dynamism can be selected for in abstract non-stationary [Bull, 2012a] and multicellular [Bull, 2013] environments.

There is growing interest in the use of GRN representations in artificial systems. RBN have been evolved to solve well-known logic tasks (e.g., [Van den Broeck & Kawai, 1990]), to model micro-array data (e.g., [Tan & Tay, 2006]), for robot maze tasks (e.g., [Preen & Bull, 2009]), etc. (see [Bull, 2012b] for an overview of evolving GRN). We are interested in the use of computational intelligence techniques as an approach to design/program unconventional computing substrates and architectures. Previous work has considered in vitro neuronal networks (e.g., [Bull, et al., 2008]), non-linear chemical media (e.g., [Toth et al., 2008]), memristors (e.g., [Howard et al., 2013]), cellular automata (e.g., [Sapin et al., 2009]), amongst others. This paper presents initial results for the design of dynamical circuits within slime mould, where each Boolean logic gate of the network is to be implemented in the living substrate, and for the design of a dynamical controller for each cell of a "smart surface" of distributed actuators and sensors.

## II. RANDOM BOOLEAN NETWORKS

Within the traditional form of RBN, a network of $R$ nodes, each with $B$ directed connections from other nodes in the network, all update synchronously based upon the current state of those $B$ nodes. Hence those $B$ nodes are seen to have a regulatory effect upon the given node, specified by the given Boolean function attributed to it. Nodes can also be self-connected. Since they have a finite number of possible states and they are deterministic, such networks eventually fall into an attractor. It is well-established that the value of $B$ affects the emergent behaviour of RBN wherein attractors typically contain an increasing number of states with increasing $B$. Three phases of behaviour were originally suggested by Kauffman through observation of the typical dynamics: ordered when $B=1$, with attractors consisting of one or a few states; chaotic when $B>3$, with a very large number of states per attractor; and, a critical regime around $1<B<4$, where similar states lie on trajectories that tend to neither diverge nor converge (see [Kauffman, 1993] for discussions of this critical regime, e.g., with respect to perturbations). Subsequent formal analysis using an annealed approximation of behaviour identified $B=2$ as the critical value of connectivity for behaviour change [Derrida & Pomeau, 1986]. Figure 1 shows the typical behaviour of networks with $R=100$ and various $B$.

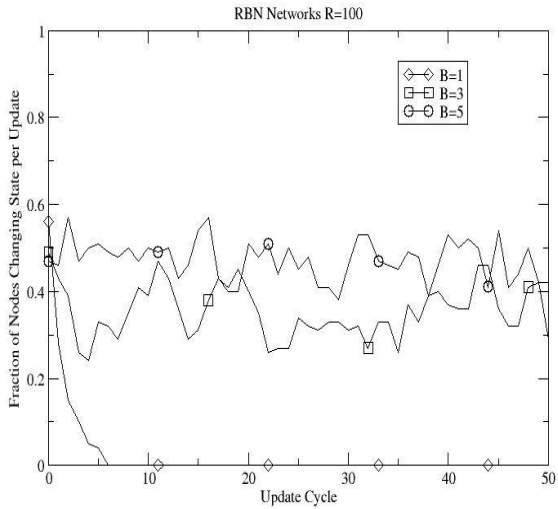

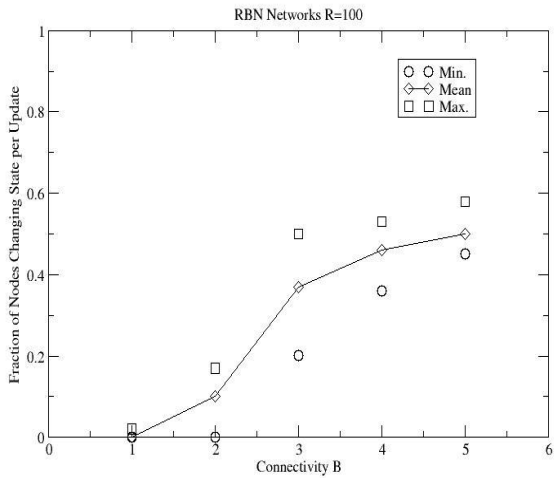

Fig. 1: Typical behavior of RBN: top, showing example temporal dynamics; and below, the average behaviour (100 runs) after 5000 update cycles.

### III. TWO MOBILE DNA MECHANISMS

In the aforementioned initial studies of mobile DNA in RBN [Bull, 2012a; 2013], structural dynamism was seen as a consequence of the actions of DNA transposons. That is, the cutting-and-pasting of segments of DNA was seen as causing a change in the connectivity structure of the GRN. Here nodes were extended to (potentially) include a second set of $B'$ connections to defined nodes. Each such dynamic nodes also performed an assigned rewiring function based upon the current state of the $B'$ nodes. Hence on each cycle, each node updates its state based upon the current state of the $B$ nodes it is connected to using the Boolean logic function assigned to it in the standard way. Then, if that node is also structurally dynamic, those $B$ connections are altered according to the current state of the $B'$ nodes it is connected to using its rewiring table. The moving of the $B$ connections of a given node via the actions/states of the $B'$ nodes is therefore seen as an abstraction of one or more of the possible effects of a mobile element as discussed above, triggered by one or more of the $B'$ nodes, causing a change in the regulatory network which affects the given node. For simplicity, the number of regulatory connections ($B$) is assumed to be the same as for rewiring ($B'$), as shown in Figure 2.

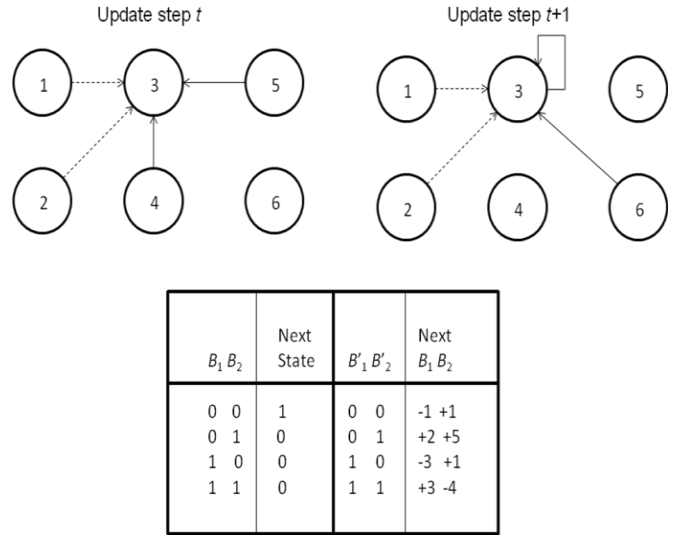

Fig. 2:. Example RBN with structural dynamism. The look-up table and connections for node 3 are shown in an $R=6$, $B=2$ network. Nodes capable of rewiring have $B'$ extra structure regulation connections into the network (dashed arrows) and use the state of those nodes to alter the standard $B$ transcription regulation connections (solid arrows) on the next update cycle ($B'=2$). Thus in the RBN shown, node 3 is a dynamic node and uses nodes 1 and 2 to determine any structural changes. At update step $t$, node 3 is shown using the states of nodes 4 and 5 to determine its state for the next cycle. Assuming all nodes are at state '0', the given node above would transit to state '1' for the next cycle and source its $B$ inputs from nodes 4-1=3 and 5+1=6 on that subsequent cycle, as defined in the first row of the table shown.

As in [Kaufman, 1993], a genetic hill-climber was considered previously, as it is here. Each RBN is represented as a list to define each node's Boolean function, $B$ connection ids, $B'$ connection ids, connection changes table entries (range +/-5), and whether it is a dynamic node or not. Mutation can therefore either (with equal probability): alter the Boolean function of a randomly chosen node; alter a randomly chosen *B* connection (used as the initial connectivity if a dynamic node); turn a node into or out of being a dynamic rewiring node; alter one of the rewiring entries in the look-up table if it is a dynamic node; or, alter a randomly chosen *B'* connection, again only if it is a dynamic node. A mutated GRN becomes the parent for the next generation if its fitness is higher than that of the original. In the case of fitness ties the number of dynamic nodes is considered, with the smaller number favoured, the decision being arbitrary upon a further tie. Hence there is a slight selective pressure against structural dynamism.

Results showed such structural dynamism was selected for under various conditions and analysis of the underlying behaviour indicated that the dynamic nodes typically experience constant rewiring during execution, usually moving between a finite set of connections as the RBN moves through its (deterministic) attractor. That is, the rewiring connections were typically made to nodes which alter their state within the attractor of the network.

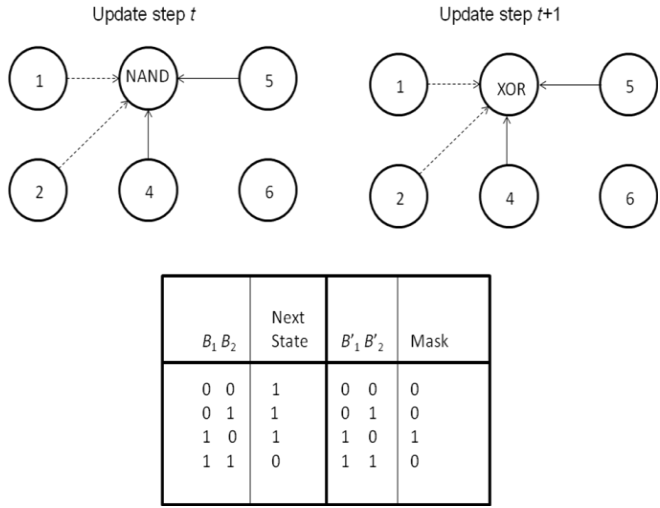

Fig. 3:. Example RBN with functional dynamism. The look-up table and connections for node 3 are shown in an *R*=6, *B*=2 network. Nodes capable of re-functioning have *B'* extra regulation connections into the network (dashed arrows) and use the state of those nodes to alter the node's Boolean logic function on the next update cycle (*B'*=*B*). Thus in the RBN shown, node 3 is a dynamic node and uses nodes 1 and 2 to determine any functional changes. At update step *t*, node 3 is shown using a NAND function to determine its state for the next cycle (encoded as 1110). Assuming all nodes are at state '0', the given node above would transit to state '1' for the next cycle and alter the first non-zero bit in its function table on that subsequent cycle, as defined in the first row of the table shown, hence changing to XOR (encoded as 0110).

Probabilistic RBN (e.g., see [Shmulevich & Dougherty, 2010]) allow for a change in node function within a given set according to a fixed distribution. It has long been noted (e.g., see [Kauffman, 1984]) that a bias in the Boolean function space of the traditional RBN - that is, a deviation from the expected average probability *P* of 0.5 for either state as the output - reduces the number of attractors and their size for a given number of nodes and connectivity. Following the node relative adjustment scheme used for connectivity, a deterministic context-sensitive form of dynamic node can be defined which incrementally alters the number of 0's or 1's in the Boolean function table for that node, as shown in Figure 3. Hence on each cycle, each node updates its state based upon the current state of the *B* nodes it is connected to using the Boolean logic function assigned to it in the standard way. Then, if that node is also functionally dynamic, the node function is altered according to the current state of the *B'* nodes it is connected to. Entries in the *B'* columns can now be either a 0 or 1. A node's Boolean logic function is stored as a binary string of $2^B$ bits. The first bit in that logic function table which is not the same as the entry in the dynamic table indexed by the current state of the *B'* connections is flipped. In this way node function can be varied in an incremental way based upon the current internal and external state of the RBN, here seen as capturing different aspects of mobile DNA than the structural dynamism. Lifetime changes are not inherited here.

## IV. PHYCHIP

We are currently designing and fabricating a massively parallel biomorphic computing device built and operated by the slime mould *Physarum polycephalum* (see [Adamatzky et al., 2012]). *Physarum polycephalum* belongs to the species of order *Physarales* and has a potentially complex life cycle. Plasmodium is a `vegetative' phase of the life cycle, wherein the slime mould exists as a single cell with a myriad of diploid nuclei. The plasmodium is visible to the naked eye and looks like an amorphous yellowish mass with networks of protoplasmic tubes. The plasmodium behaves and moves as a giant amoeba and feeds on bacteria, spores and other microbial creatures and micro-particles. When foraging for its food the plasmodium propagates towards sources of food particles, surrounds them, secretes enzymes and digests the food. When several sources of nutrients are scattered in the plasmodium's range, the plasmodium forms a network of protoplasmic tubes connecting the masses of protoplasm at the food sources (Figure 4).

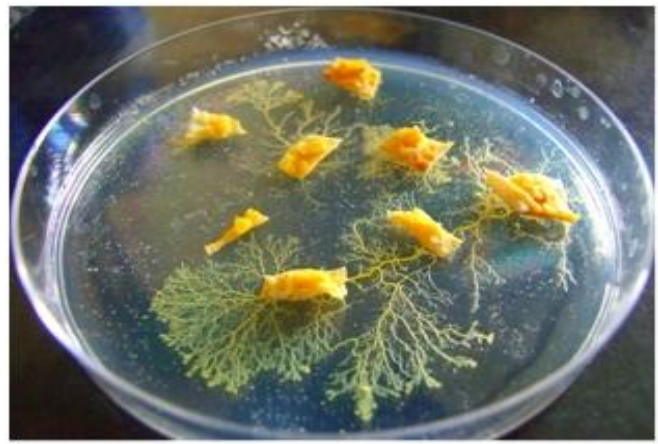

Fig. 4: Showing the slime mould *Physarum polycephalum* in the plasmodium phase, attracted to eight spatially separated food sources. From [Adamatzky et al., 2012].

Due to its unique features and relative ease of use for experimentation, the plasmodium has become a test biological substrate for the implementation of various computational tasks. The induction of behaviour and spatial form/pattern is determined partly by the environment, specifically nutrient quality and substrate hardness, dryness etc. *Physarum* can therefore be viewed as a computational material based upon the modification of protoplasm transport via the presence/absence of external stimuli. Moreover, it is sensitive to illumination and AC electric fields and therefore allows for the parallel and non-destructive input of information. *Physarum* is typically used such that it represents results of computation by the configuration of its body. The problems solved by the plasmodium include mazes, calculation of efficient networks, construction of logical gates, data clustering, and robot control (see [Adamatzky, 2010] for an overview).

The *Physarum*-based biomorphic device – "Phychip" - envisaged is shown in Figure 5. Protoplasmic tubes of *Physarum* coated with conductive substances interfaced with living blobs of plasmodium are the basic units. The living blobs play the role of sensors and processing units. Such blobs will communicate with each other using fast electrical signal transfer via conductor coated tubes, and slow electrical and bio-chemical signal transfer along living protoplasmic tubes.

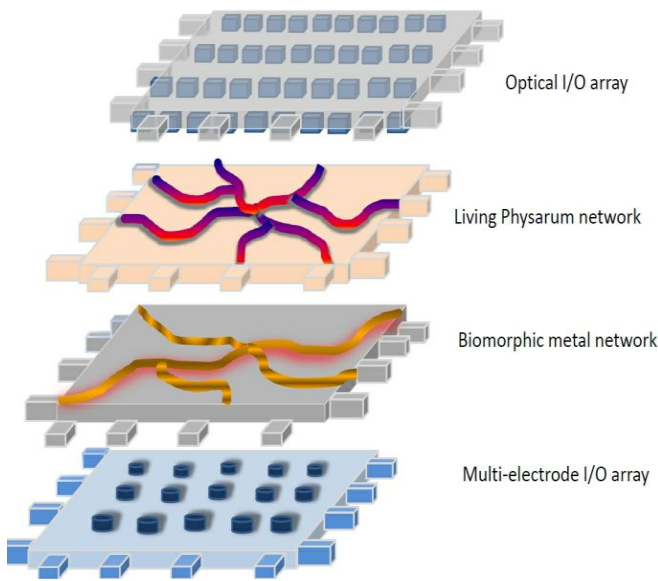

Fig. 5: Proposed scheme of the *Physarum* chip. The chip will consist of interwoven living and conductor-coated networks and layers of optical and electrical I/O. From [Adamatzky et al., 2012].

In terms of classical computing architectures, the following characteristics can be attributed to such chips:-

- massive parallelism: there are thousands of elementary processing units, micro-volumes, in a slime mould colonised in a Petri dish
- local connectivity: micro-volumes of cytoplasm change their states, due to diffusion and reaction, depending on states of, or concentrations of, reactants, shape and electrical charges in their closest neighbours
- parallel input and output: *Physarum* computes by changing its shape, can record computation optically; *Physarum* is light sensitive, data can be inputted by localized illumination
- fault tolerance: being constantly in a shape changing state, *Physarum* chips can restore their architecture even after a substantial part of a protoplasmic network is removed.

One planned use of the chip is to implement RBN within the network of plasmodium blobs, using computational intelligence techniques to determine the design for a given task. As noted above, it has already been shown how some Boolean logic gates can be implemented in the slime mould, e.g., Tsuda et al. [2004] describe the creation of AND, OR and NOT gates. Hence the design of RBN using the restricted set of *B*-input {AND, NAND, OR, NOR} gates and spatially local connectivity (nodes can only connect to the eight nodes surrounding them, or fewer on edges/corners) is considered here.

## V. PHYCHIP EXPERIMENTATION

In the following, two well-known logic problems are used to begin to explore the characteristics and capabilities of the general approach. The multiplexer task is used since they can be used to build many other logic circuits, including larger multiplexers. These Boolean functions are defined for binary strings of length $l = k + 2^k$ under which the $k$ bits index into the remaining $2^k$ bits, returning the value of the indexed bit. Hence the multiplexer has multiple inputs and a single output. Adders have multiple inputs and multiple outputs. As such, a simple example is used here. A simple sequential logic task is also used here - the JK latch. In all cases, the correct response to a given input results in a quality increment of 1, with all possible binary inputs being presented per solution evaluation. Upon each presentation of an input, each bit is applied to the first connection of each corresponding node in the RBN. The RBN is then executed for 10 cycles. The value on the predetermined output node(s) is then taken as the response. All results presented are the average of 20 runs. Experience found $R=5 \times 5= 25$ nodes/blobs was useful across the problems explored here.

Figure 6 shows performance on $k=2$ versions of the three tasks: the 6-bit multiplexer (opt. 64), 2-bit adder (opt. 16), and 2-input JK latch (opt. 4). Only structural dynamism is used given the non-trivial physical manipulation of the plasmodium needed to exhibit a given logic gate [Tsuda et al., 2004]. Given the known underlying dynamics of RBN (section II), $B=2$. As can be seen, optimal performance is reached in all cases, with varying numbers of dynamic nodes. That is, discrete dynamical circuits capable of the given logic functions have been designed for potential implementation on the Phychip. Figure 6 also shows the performance of the equivalent traditional RBN, i.e., without structural dynamism, which is statistically significantly worse on the multiplexer and adder (T-test, $p<0.05$), and the same on the latch (T-test, $p \geq 0.05$).

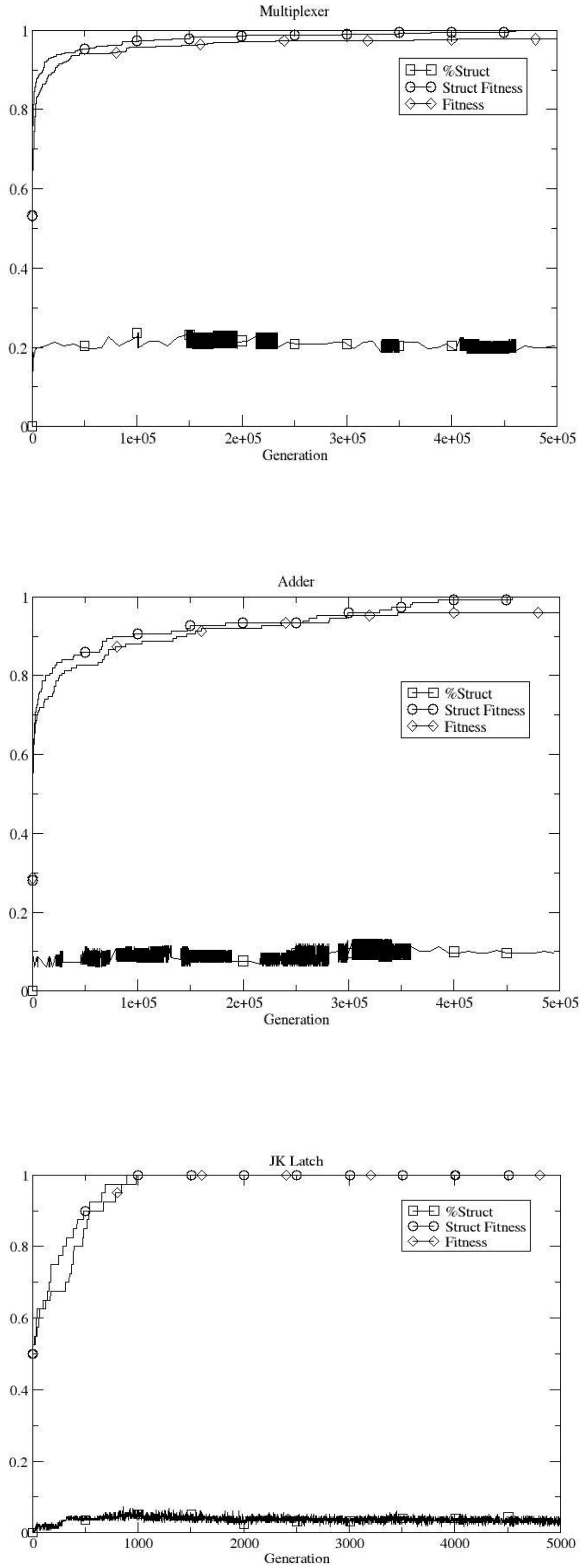

Fig. 6: Typical evolutionary performance of RBN with and without mobile DNA-inspired mechanisms for the four benchmark tasks. Fitness shown as fraction of correct inputs.

## VI. SMART SURFACE

On the surface of many cells are thin hair-like structures; cilia. Once considered vestigial, these cilia are functioning organelles of which there are two types. Non-motile or primary cilia, typically serve as sensory organelles with roles in chemical sensation (as in olfaction), signal transduction (rod photoreceptors in vision) and control of cell growth. Motile cilia are often found in clusters and move in coordinated ways, capable of registering surrounding fluid flow such as in the trachea or kidneys. If the combined properties of these cilia could be created and controlled, they could perform the functions of both motion and sensory perception in an intelligent system, sensing and identifying properties of objects on the artificial cilia surface, and potentially moving in a coordinated way to project the object along a predetermined trajectory according to its properties. In effect, a multitude of biologically inspired cilia could create the emergent properties of both sorting and transporting objects around an intelligent manipulator surface. We are currently building an intelligent autonomous massively parallel manipulator with the aim of achieving the distributed sensing, recognition, analysis, sorting, transportation and manipulation of light-weight objects. Some of the potential benefits of the decentralised manipulator surface are:

- absence of a central processor responsible for all computation - all cells of the surface act in parallel, fulfilling a collective task
- the number of simultaneously manipulated objects is not restricted by computational capability of the central processor
- the system is scalable, i.e. manipulator area can be varied (by adding/removing cells) without reprogramming.
- the system is robust against individual component faults - decentralised control and parallel operation ensure that no single component fault results in the overall system faulting.

Previous approaches to implement such systems include airjets (e.g., [Moon & Luntz, 2006]), mechanical wheel-based arrangements (e.g., [Murphey & Burdick, 2004]), and in micro electro mechanical systems (MEMS) (e.g., [Ataka et al., 2009]). The hardware implementation of our system will be based on an array of piezo-actuators (Figure 7). This system will serve as an experimental platform and enable us to demonstrate many of the properties listed above. It is expected that each axis of each actuator will consist of eight piezo elements; object movement will be achieved by driving the appropriate transducer element groups. Mass induced changes in piezo-electric element resonant frequency established through monitoring impedance variation will be used to sense the presence of an object.

The design of distributed controllers for such systems is non-trivial and hence computational intelligence techniques may be usefully employed. In the only known related work,

Matignon et al. (e.g., [2010]) have explored using reinforcement learning approaches. As noted above, we are interested in the evolution of an RBN controller to exist within each cell of the surface, sensing and acting locally, i.e., akin to the cells of multicellular organisms (see [Bull, 2012b] for a review of related multi-GRN work). In the aforementioned initial work on structural dynamism in RBN it was found that the new mechanism can facilitate behavioural differentiation in two coupled cells [Bull, 2013]. To act as a controller, each RBN receives five sensor inputs as to the presence or absence of an object: itself and the four cardinal direction cells (von Neumann neighbourhood of cellular automata). To facilitate the emergence of useful cell-cell communication, each RBN also receives an input from a predetermined node in each of the RBN of its four cardinal neighbouring cells. Thus, following the input scheme above, the first nine nodes of a given RBN each receives an external input as its first connection. Possible outputs are to move the actuator in one of the four cardinal directions, hence two output nodes are predetermined.

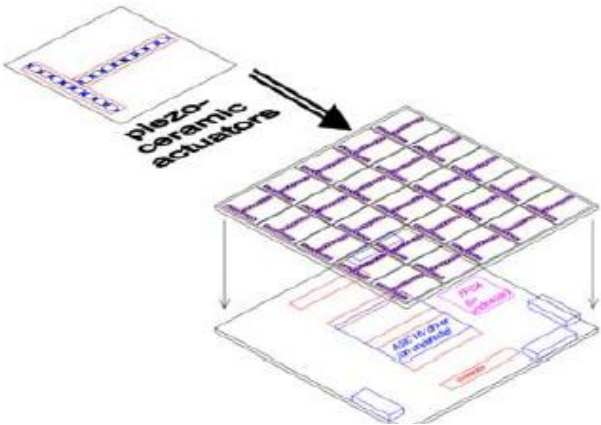

Fig. 7: Proposed piezo-ceramic actuator layout of each cell in a multi-celled smart surface.

### VII. SMART SURFACE EXPERIMENTATION

In the following experiments we begin to explore a fundamental aspect of controller design for such distributed systems – the ability to correctly recognize an object and respond appropriately. More specifically, the task of distinguishing between two objects by moving them in different directions is considered. In these initial experiments, a box of 3x1 cells must be moved as far north as possible from the middle of a surface of 12x12 cells having been placed horizontally. That is, the object is placed on cells 75, 76 and 77 of the 144, with the fitness calculated as the distance between the middle section of the object and cell 76. The surface is then reset, i.e., each RBN is set to its start configuration, and a box of 5x1 cells is similarly placed horizontally in the middle of the surface and must be moved as far south as possible. In each case, each RBN updates in parallel internally for 10 cycles for a given sensor input before its action is determined. This process is repeated 10 times. If all cells under an object give the same action, the object moves in that direction. Fitness in each case is the distance of the middle of the object from the middle of the surface, with the fitness of an RBN simply the sum of the fitness in each of the two scenarios. Here $R=20$ and $B=2$, all Boolean functions are allowed, results averaged over 20 runs.

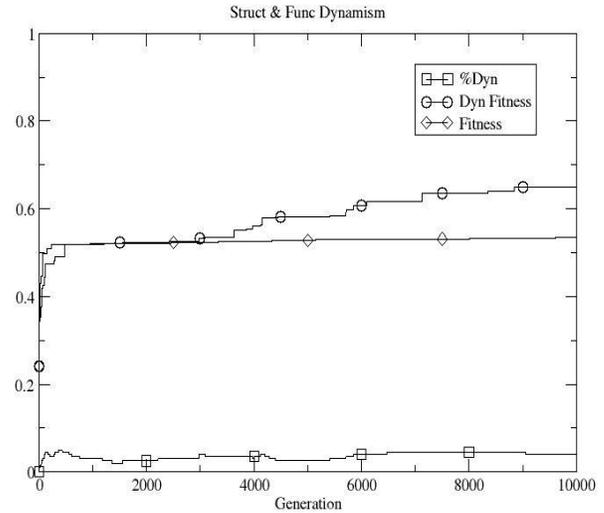

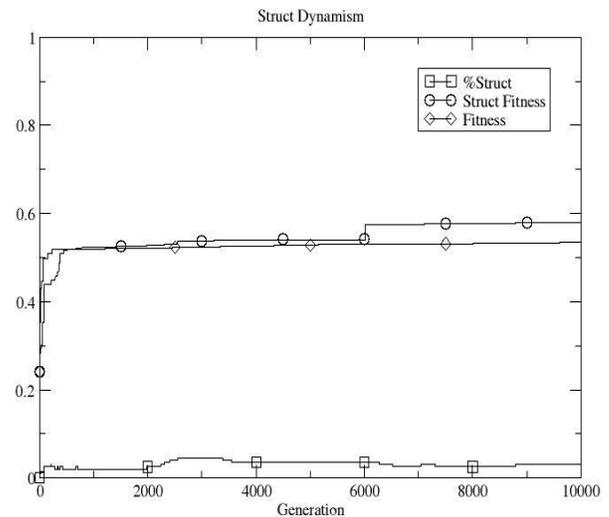

Fig. 8: Typical evolutionary performance of RBN with and without mobile DNA-inspired mechanisms for the benchmark surface task. Fitness shown as fraction of known optimum.

Figure 8 shows performance with structural dynamism (as above) and both structural and functional dynamism. As can be seen, in both cases typical performance is better than the standard RBN representation (T-test, $p<0.05$). Moreover, the combined dynamism performs better than just the structural version (T-test, $p<0.05$). Both versions of the RBN using the mobile DNA-inspired mechanisms evolve to significantly separate the two objects whereas the traditional approach typically only evolves to move both objects in the same direction. That is, only with the extra mechanisms are the cells in the ambiguous positions under the objects able to distinguish between the two.

VIII. CONCLUSION

There is a growing body of work within computational intelligence which explores representations more closely analogous to the genetic machinery seen in nature, i.e., artificial regulatory networks. Adoption of these relatively generic representations creates the opportunity to exploit new mechanisms from microbiology. That is, molecular biologists have identified a variety of mechanisms through which changes in DNA occur in natural regulatory networks in ways other than the processes which inspired the traditional heuristics of evolutionary computation: specific biochemical processes generate novelty through targeted DNA restructuring based upon the internal and external state of a GRN *during* the organismal lifecycle. This paper has presented initial results from using two new mechanisms within GRN to solve computational tasks. Current work is seeking to incorporate other mechanisms, as well as determining the general effectiveness of such systems.

This work was partially supported by EPSRC grant no. EP/H023631/1 and EU FP7-ICT project no. 316366.